\documentclass[preprint,5p, twocolumn,number,floatfix]{elsarticle}
\usepackage{hyperref}
\usepackage{amssymb,amsmath}
\usepackage[usenames,dvipsnames]{color}
\usepackage{soul}
\usepackage{pifont}
\usepackage{epsfig}
\usepackage{stix}
\usepackage{longtable}
\usepackage{array,multirow}
\begin{document}
\begin{frontmatter}
\title{The effects of inhomogeneities on scroll-wave dynamics in an anatomically 
realistic mathematical model for canine ventricular tissue}
\author[1]{K. V. Rajany\corref{cor1}\fnref{fn1}}
        \ead{rajanyk@iisc.ac.in}
\address[1]{Centre for Condensed Matter Theory, Department of Physics, Indian
Institute of Science, Bangalore 560012, India. }
\cortext[cor1]{Corresponding author}
\fntext[fn1]{Present address:S.A.R.B.T.M. Govt. College, Koyilandy, Calicut, India}
\author[2]{Rupamanjari Majumder}
\address[2]{Department of Fluid Dynamics, Pattern Formation and Biocomplexity,
Max Planck Institute for Dynamics and Self-Organization.}
\author[3]{Alok Ranjan Nayak}
\address[3]{International Institute of Information Technology, Bhubaneswar.}
        \author[4]{Rahul Pandit\fnref{fn3}}
        \ead{rahul@iisc.ac.in}
        \fntext[fn3]{also at Jawaharlal Nehru Centre For
Advanced Scientific Research, Jakkur, Bangalore, India.}
\address[4]{Centre for Condensed Matter Theory, Department of Physics, Indian
Institute of Science, Bangalore 560012, India. }
\date{\today}
\begin{abstract}

Ventricular tachycardia (VT) and ventricular fibrillation (VF) are lethal
rhythm disorders, which are associated with the occurrence of abnormal
electrical scroll waves in the heart. Given the technical limitations
of imaging and probing, the \textit{in situ} visualisation of these
waves inside cardiac tissue remains a challenge.  Therefore, we must,
perforce, rely on \textit{in silico} simulations of scroll waves in
mathematical models for cardiac tissue to develop an understanding of
the dynamics of these waves in mammalian hearts. We use direct
numerical simulations of the Hund-Rudy-Dynamic (HRD) model, for canine
ventricular tissue, to examine the interplay between electrical
scroll-waves and \textit{conduction} and \textit{ionic}
inhomogeneities, in anatomically realistic canine ventricular
geometries with muscle-fiber architecture.  We find that
millimeter-sized, distributed, conduction inhomogeneities cause a
substantial decrease in the scroll wavelength, thereby increasing the
probability for wave breaks; by contrast, single, localized,
medium-sized ($\simeq$ cm) conduction inhomogeneities, exhibit the
potential to suppress wave breaks or enable the self-organization of
wave fragments into stable, intact scrolls. We show that ionic
inhomogeneities, both distributed or localised,  suppress scroll-wave
break up.  The dynamics of a stable rotating wave is not affected
significantly by such inhomogeneities, except at high concentrations of
distributed inhomogeneities, which can cause a partial break up of
scroll waves.  Our results indicate that inhomogeneities in the canine
ventricular tissue are less arrhythmogenic than inhomogeneities in
porcine ventricular tissue, for which an earlier \textit{in silico}
study~\cite{rpm-inhomo} has shown that the inhomogeneity-induced
suppression of scroll waves is a rare occurrence.
\end{abstract}
\end{frontmatter}
\graphicspath{{./final_figures/}}
\section{Introduction}

Life-threatening cardiac arrhythmias, such as ventricular tachycardia (VT) and
ventricular fibrillation (VF), which are the leading cause of death in the
industrialised world, are associated with unbroken (for VT) or broken (for VF)
spiral or scroll waves of electrical activation in ventricular
tissue~\cite{davidenko93,clayton08,clayton11,trayanova11,cherry08,sneyd,reviewroyal,review1,rpm1,shajahan1}.
The mechanisms leading to the formation of such scroll waves, their dynamics,
and their interactions with other structural features in mammalian hearts
continue to be an active area of
research~\cite{rpm1,shajahan1,rpm2,rpm3,alok2,alok3,ikeda,limmaskara,shajahan2,shajahan3},
for many different features can be arrhythmogenic. The heart has a complex,
inhomogeneous structure and intricate geometry; it includes conducting tissue,
with cardiac myocytes, nonconducting tissue, with fibroblasts, muscle fibers, a
sheet structure, blood vessels, and scar tissues. The elucidation of the
interactions of scroll waves of electrical activation with these multi-scale
structures in cardiac tissue is required to develop (a) a detailed
understanding of VT and VF and (b) possible defibrillation strategies, either
pharmacological or electrical. 
 
Experimental and numerical studies have shown that heterogeneities in cardiac
tissue affect the dynamics of spiral or scroll waves and their break up (see,
e.g., Refs.~\cite{rpm1,shajahan1,rpm2,rpm3,alok2,alok3,shajahan2,shajahan3} and
references therein), and thereby, both atrial and ventricular fibrillation.
However, experiments on intact mammalian hearts are usually laborious,
time-consuming, expensive, and still hampered by the considerable challenges in
the visualization of electrical activity, below the tissue surface.
~\cite{rpm1,shajahan1,rpm2,rpm3,alok2,alok3,ikeda,limmaskara,shajahan2,shajahan3}.
Up until now, the experimental visualization of scroll filaments in the intact
heart was only possible during fibrillation; a recent study~\cite{christoph}
has used high-resolution, four-dimensional (4D), ultrasound-based imaging to
map the occurrence of mechanical filamentous phase singularities that coexist
with electrical phase singularities during cardiac fibrillation; 4D cardiac
electromechanical activation imaging has also been used in
Ref.~\cite{Grondin19}.
Thus, numerical simulations or \textit{in silico} studies, which use
mathematical models for cardiac tissue, have become an increasingly promising
alternative to \textit{in vivo} and other experimental  studies of scroll-wave
dynamics in cardiac tissue.

\textit{In silico} studies also give us great flexibility in changing
individual parameter values (to explore, e.g., the effects of different
medications on various ion-channel properties). However, such \textit{in
silico} studies are computationally challenging, especially if we use (a)
physiologically realistic mathematical models, (b) anatomically realistic
geometries, and (c) structural complexities like muscle-fiber orientation.  It
is not surprising, therefore, that many numerical studies use either simplified
cardiac-tissue mathematical models or idealised geometries, e.g., a cubical
domain or a wedge~\cite{reviewroyal}. 

For reviews of the dynamics of scroll waves in the presence of structural and
electrophysiological heterogeneities in cardiac tissue, we refer the reader to
Ref.~\cite{clayton08,clayton11,trayanova11,cherry08,review1}. There have been
studies of fiber-rotation-induced scroll-wave turbulence (see, e.g.,
Refs.~\cite{fentonkarma1, fentonkarma2, rupamanjariF} and references therein);
in particular,  they have investigated the effects of the anisotropy,
because of fiber and sheet orientation, on such turbulence. Numerical studies,
which use simplified models, have shown that an inhomogeneous excitable
three-dimensional (3D) system, with a gradient of excitability, promotes
twisting of the scroll waves, and can cause a simple scroll wave to change to a
meandering scroll wave~\cite{yanggao}. Thin, rod-like heterogeneities have been
shown to suppress otherwise developing spatiotemporal chaos; furthermore, (a)
they clear out already existing chaotic excitation patterns~\cite{spreckelson}
and (b) spiral waves are either attracted towards or repelled from the centers
of such inhomogeneities~\cite{Pawelsandersetal}. Computer-simulation studies
have studied the effects of the structure of the heart on scroll waves in
mathematical models~\cite{clayton08,clayton11,trayanova11,cherry08}; for recent
\textit{in silico} studies of mathematical models for different mammalian
hearts see, e.g., Refs.~\cite{rpm-inhomo,G19,Zimik20}, for porcine- rabbit-,
and human-ventricular models, respectively.

\begin{figure}
\begin{center}
\includegraphics[width=0.9\linewidth]{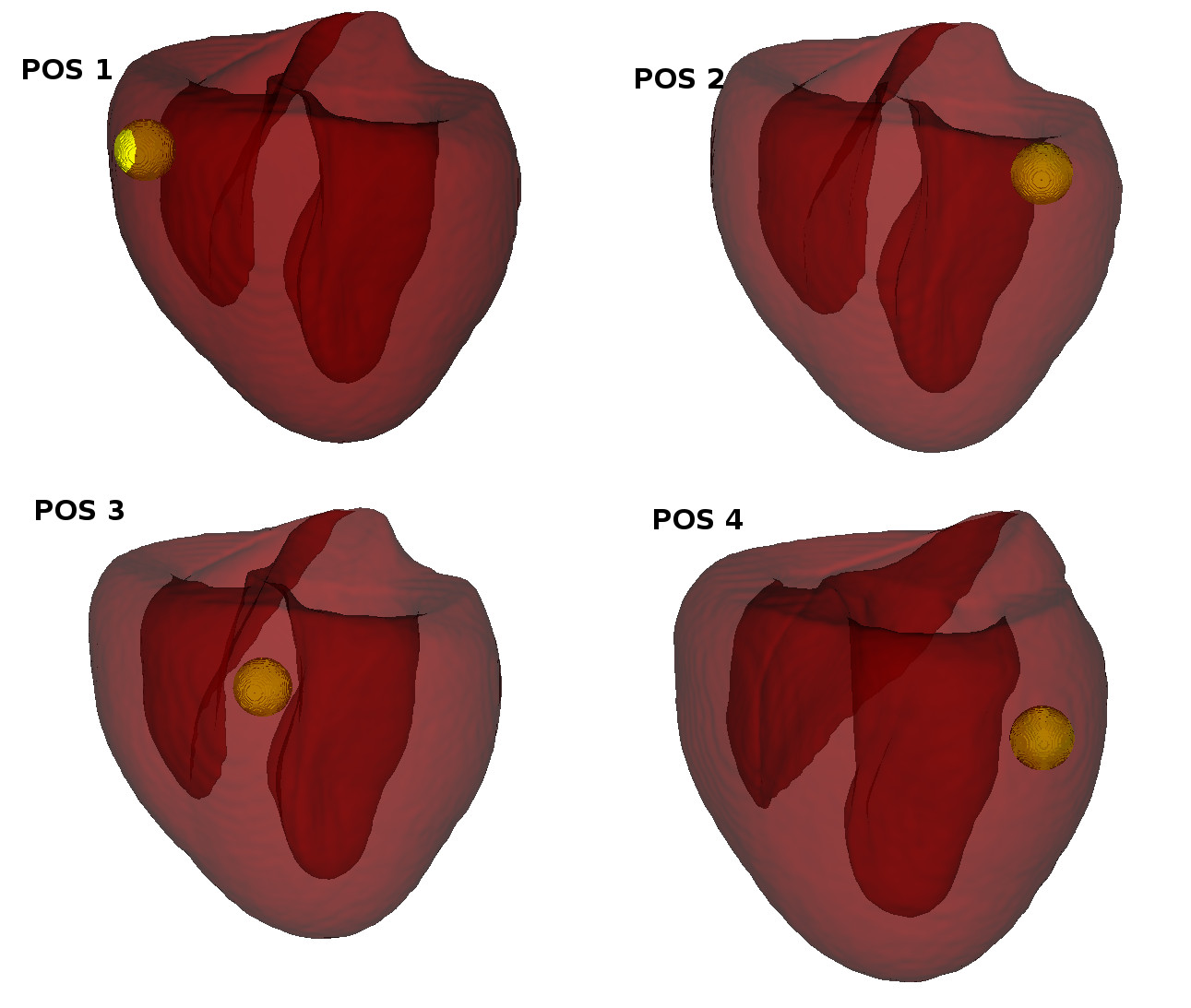}
\caption {The four illustrative positions of the localized, spherical inhomogeneities,
inside the ventricular geometry, that we use for our simulations: One
on the surface, partly inside and partly outside the tissue; one well
inside the ventricular wall; one in the septum; we choose the last position,
such that the scroll-wave filament comes close to the
inhomogeneity during our numerical simulation.}
\label{spher_inhomo_pos}
\end{center}
\end{figure}

We investigate the interaction of scroll waves with two types of
inhomogeneities in a realistic heart geometry for canine ventricles, conduction
inhomogeneities and ionic inhomogeneities. As a person ages, the concentration
of nonconducting tissue increases; also, scars or dead tissue act as conduction
blocks. These are the physiological equivalents of distributed conduction
inhomogeneities and localized conduction inhomogeneities, respectively.
Changes in different ionic conductivities do occur in mammalian hearts, in
abnormal situations, or because of medicines or other external factors; these
changes can be modelled as ionic inhomogeneities.  Majumder, \textit{et al.}
have conducted a study of scroll-wave dynamics in a mathematical model for a
pig heart with inhomogeneities~\cite{rpm-inhomo}. In particular, they have
studied the case in which the ventricle initially has a single rotating scroll
wave; and they have then investigated how this single scroll evolves, as a
result of the inhomogeneity that is introduced. We go beyond the study of
Ref.~\cite{rpm-inhomo} by including initial conditions that produce meandering
as well as broken scroll waves, in the absence of inhomogeneities ;we then let
these waves evolve in the presence of inhomogeneities.  Furthermore, we use a
mathematical model for a dog heart instead of a pig heart. We compare the
dynamics of scroll waves in these mathematical models for two different
mammalian species, which have hearts of comparable sizes, but different
electrophysiology. We find that inhomogeneities in the canine
ventricular tissue are less arrhythmogenic than inhomogeneities in
porcine ventricular tissue, for which an earlier \textit{in silico}
study~\cite{rpm-inhomo} has shown that the inhomogeneity-induced
suppression of scroll waves is a rare occurrence.

The remaining part of this paper is organised as follows. In
Sec.~\ref{sec:ModMeth} we present the model we use and the numerical methods we
employ for our \textit{in silico} study of this model.  We discuss our results,
for both conduction- and ionic-type inhomogeneities, in Sec.~\ref{sec:Results}.
We end with conclusions and a discussion of our results in
Sec.~\ref{sec:Conclusions}.

\begin{figure}
\begin{center}
\includegraphics[width=0.9\linewidth]{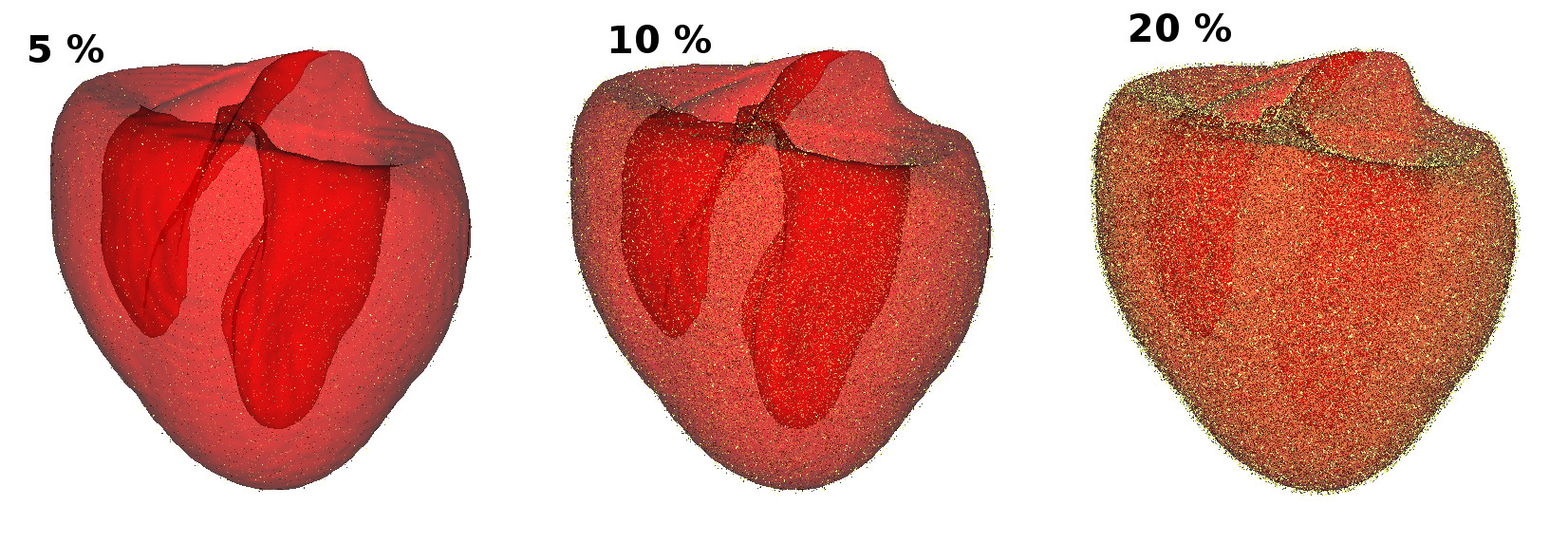}
\caption
[The distribution of small-size inhomogeneities, inside the ventricular
geometry that we use for our simulations]
{The distribution of small-size inhomogeneities, inside the ventricular 
geometry that we use for our simulations: we employ three different concentrations
(see text), namely, 5\%, 10\%, and 20\%.}
\label{distr_inhomo_pos}
\end{center}
\end{figure}
\section{Models and Methods}
\label{sec:ModMeth}

For the canine-ventricular geometry that we use in our computations, we have
downloaded diffusion-tensor magnetic-resonance-imaging (DTMRI) data
from~\cite{cmiss-site}; and we have then processed these data to produce our
simulation geometry, with the locations of heart points and fiber directions.

We use the electrophysiologically detailed Hund-Rudy-Dynamic (HRD)
model~\cite{hundrudy} for canine-ventricular myocytes to describe the dynamics
at the cell level. The HRD model has $45$ variables, including membrane and
intercellular currents, gating variables, ionic concentrations, and the
transmembrane potential $V$. We give a detailed description of this model in
the Supplementary Material~\cite{appndx-supp}. We use the monodomain approach for
ventricular tissue; and the spatiotemporal evolution of the transmembrane
potential $V$ is given by the following partial-differential equation (PDE),
which is of the reaction-diffusion type (all the other variables follow coupled
ordinary differential equations (ODEs)):
\begin{equation}
\label{monodomain}
\frac{\partial V}{\partial t} = \nabla .\mathcal{D} \nabla V -\frac {I_{ion}+I_{applied}}{C_m} , 
\end{equation}
where $I_{ion}$ contains all the ionic currents, $I_{applied}$ is the external
current, $C_m$ the capacitance density, and $\mathcal{D}$ the diffusion tensor
that accounts for the conductivity of the tissue and propagation between cells;
its elements are~\cite{tentusscher}:
\begin{equation}
\mathcal{D}_{ij} = D_T * \delta_{ij} + (D_L - D_T)\alpha_i \alpha_j,
\end{equation}
where $\alpha$ is the vector describing the muscle-fiber direction, and
$D_L$ and $ D_T $ are the conductivities in the longitudinal
and transverse fiber directions, respectively.

We impose Neumann boundary conditions by using the phase-field approach of
Ref.~\cite{phasefield} (see the Supplementary Material~\cite{appndx-supp}). We
solve the ODEs via the forward-Euler scheme and the PDEs by using a
central-difference finite-difference method. We use a spatial resolution of
$0.02$ cm and a time step of $0.01$ms for our simulations. Such simulations are
computationally challenging. To meet this challenge, we have developed an
efficient CUDA code that we run on computer clusters that use graphics
processing units (GPUs). This requires specialised
programming~\cite{RajanyThesis2020}.  We have checked our numerical results by
comparing them with results for various ionic currents and the conduction
velocity in the original HRD-model paper~\cite{hundrudy}. We have also carried
out trial runs for a long time ($\simeq 10$s) to ensure the numerical stability
of our results.

We consider the following two types of inhomogeneities in our HRD
ventricular-tissue model: (a) conduction inhomogeneities because of a change in
the diffusivity of the medium; we simulate these by setting the diffusion
constant $D=0$ at the locations of the inhomogeneities; (b) ionic
inhomogeneities that arise from a change in any of the ionic properties across
the cell membrane; in particular, we study ionic inhomogeneities by varying the
L-type Calcium-ion-channel current $I_{CaL}$ by reducing the value of the
relevant parameter to $\gamma_{Cao}=0.0314$.  We have conducted extensive
numerical studies with (a) medium-sized spherical inhomogeneities, located at
four representative positions, as shown in Fig.~\ref{spher_inhomo_pos}, and (b)
small-sized inhomogeneities, distributed throughout the simulation domain, with
concentrations 5\%, 10\% and 20\% of the whole volume of the tissue, as shown
in Fig.~\ref{distr_inhomo_pos}.

\begin{figure}
\includegraphics[width=0.9\linewidth]{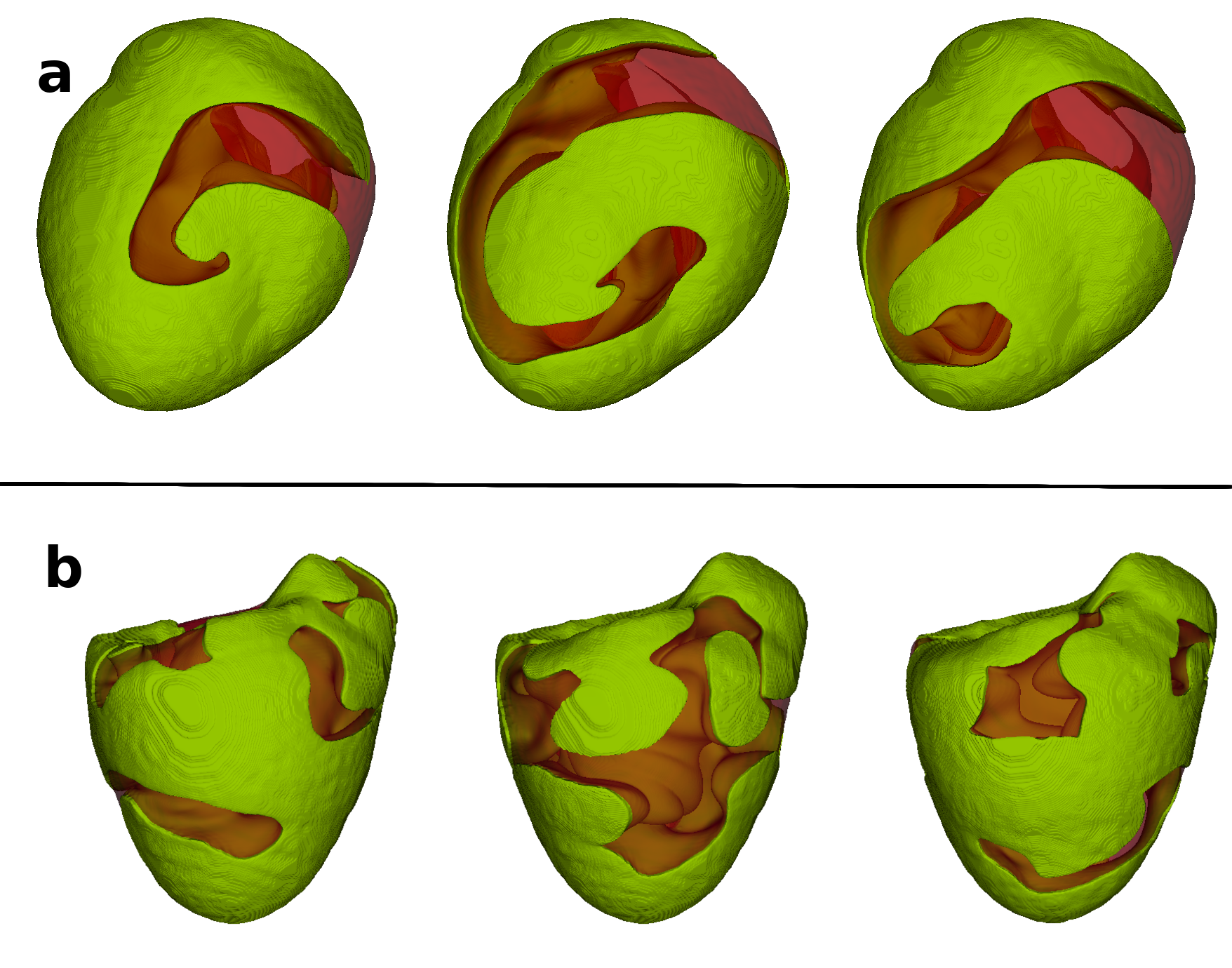}
\caption[Spatiotemporal evolution of scroll waves in the HRD model in the
anatomically realistic geometry for a canine ventricle without any
inhomogeneities]
{Two-level isosurface plots of the transmembrane potential $V$ illustrating the 
spatio-temporal evolution of scroll waves in the HRD model, in the 
anatomically realistic geometry for a canine ventricle (see text) and without any 
inhomogeneities. We use two different parameter sets: the first yields a rotating 
scroll wave (upper panel) and the second broken scroll waves interacting with each 
other (lower panel). For the full spatio-temporal evolution of $V$ see the videos
S41 (for the upper panel) and S42 (for the lower panel) in the Supplementary Material~\cite{chapt4movies}.}
\label{noinhomo_dyn}
\end{figure}

With no inhomogeneity in our simulation domain, we have two types of
scroll-wave dynamics: In the first case, there is a single, rotating,
meandering scroll ($G_{Kr}=5\times0.0138542$ and $\gamma_{Cao}=3\times0.3410$);
in the second, the scroll breaks up and its pieces interact with each other
($G_{Kr}=3\times0.0138542$ and $\gamma_{Cao}=1\times0.3410$)~\cite{rajanyhrd}.
We illustrate these two types of scroll-wave dynamics, without any
inhomogeneities, in Fig.~\ref{noinhomo_dyn}. We now introduce inhomogeneities
into our simulation domain to investigate how they affect the dynamics of these
scroll waves.  We study $28$ cases in total. For each one of these cases we run
our simulation for a time that is equivalent to $2.5$s in real time.

\begin{table*}[]
\centering
\caption[Summary of our results for the effects of conduction-type
inhomogeneities on scroll-wave dynamics]{This table summarizes our results for the effects of conduction-type 
inhomogeneities on scroll-wave dynamics.}
\begin{tabular}{|c|p{2.5cm}|p{5cm}|p{2cm}|p{5cm}|}
\hline
&\multicolumn{2}{c}{localized spherical inhomogeneity} & \multicolumn{2}{|c|}{distributed small-scale inhomogeneities}\\
\cline{2-5}
\raisebox{-3\normalbaselineskip}[0pt][0pt]{\multirow{4}{*}{\rotatebox[origin=l]{90}{Broken scroll}}}

&position1 & broken scrolls move away from the inhomogeneity and eventually move out of the domain. &{\textbf {5 \%}} & rotating scroll with reduced wavelength. \\
& position2 & breakup partially suppressed; broken scrolls change to a rotating scroll; it breaks again because of functional instability; and the cycle repeats . &{\textbf {10 \%} } & wavelength reduces and break up occurs. \\ 
&position3 & broken scrolls continue to break up and interact; scroll filament moves away from the inhomogeneity; breaking is not suppressed.& {\textbf {20 \% }} & wavelength 
reduces significantly and break up is considerable.\\
&position4 & breaking suppressed in the long run; and broken scrolls change to a
rotating scroll. 
	
 & & \\ 
\hline
 \raisebox{-1\normalbaselineskip}[0pt][0pt]{\multirow{4}{*}{\rotatebox[origin=b]{90}{Rotating scroll}}} 
&position1 & continues rotating&{\textbf {5 \%}} & wavelength reduces; and the scroll
wave continues to rotate. \\
&position2 & continues rotating\newline & {\textbf {10 \%} } & wavelength reduces and 
the scroll breaks up. \\
&position3 & continues rotating\newline & {\textbf {20 \% }} & wavelength reduces by a
large amount and there is considerable break up. \\
&position4 & scroll continues rotating. \newline & & \\ 
\hline
\end{tabular}
\label{table1}
\end{table*}
\section{Results}
\label{sec:Results}

In this Section, we present the results of our \textit{in silico} studies. We
first present the results of our simulations with conduction-type
inhomogeneities, and then the results we have obtained with ionic-type
inhomogeneities. 

\subsection{Conduction Inhomogeneity: Effects on Scroll-wave Dynamics}
\label{subsec:CondInhom}

Before we describe our results in detail, we refer the reader to
Table~\ref{table1}, which summarizes our principal results for the effects of
conduction-type inhomogeneities on scroll-wave dynamics. 

\begin{figure}
\begin{center}
\includegraphics[width=0.7\linewidth]{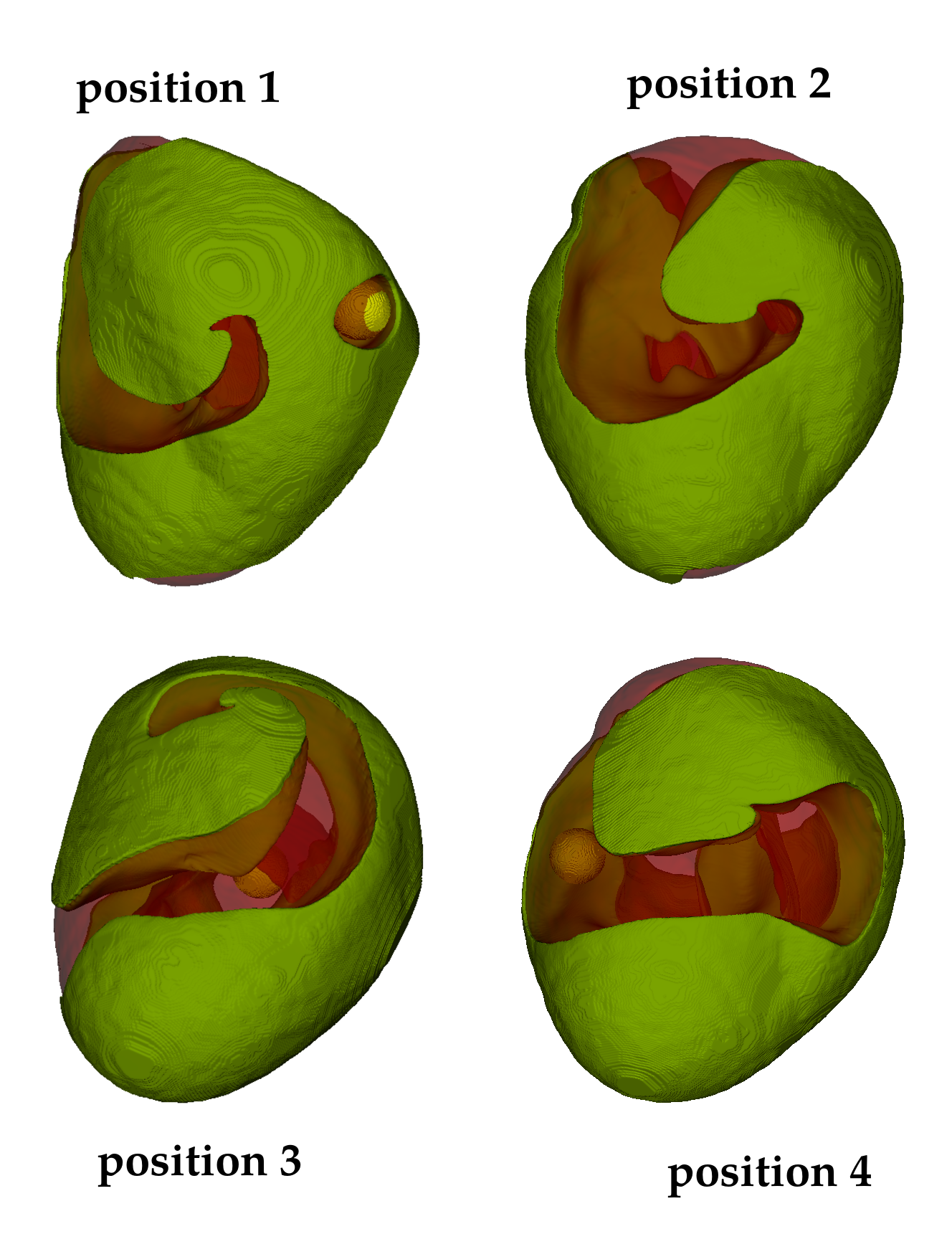}
\caption[Spatiotemporal evolution
of scroll waves in the HRD model in the anatomically realistic canine
ventricular geometry, with a medium-sized spherical inhomogeneity of
conduction type placed at four different positions and with the initial configuration
that produces a rotating scroll]
{Two-level isosurface plots of the transmembrane potential $V$ illustrating 
scroll-wave dynamics, with a medium-sized spherical inhomogeneity of
conduction type placed at position $1, \, 2, \, 3,$ and $4$ in the domain 
(see Fig.~\ref{spher_inhomo_pos}). The initial configuration is obtained
from the parameter set that produces a rotating scroll wave without the
inhomogeneity; we depict the final 
configurations here.  The rotating scroll is unaffected by the inhomogeneity.
For the full spatio-temporal evolution of $V$ see the videos S47 , S48
, S49  and S410 in the Supplementary Material~\cite{chapt4movies}. }
\label{cond_spher_1r}
\end{center}
\end{figure}

Figure~\ref{cond_spher_1r} shows the final configuration of the evolution of
scroll waves with a localized, spherical inhomogeneity (obstacle) of conduction
type, for the four different, representative positions, on an initially
rotating-scroll configuration. In this case we do not observe a qualitative
change in the scroll-wave dynamics introduced by the inhomogeneity. The
rotating scroll continues to rotate and is stable.

\begin{figure}
\begin{center}
\includegraphics[width=0.95\linewidth]{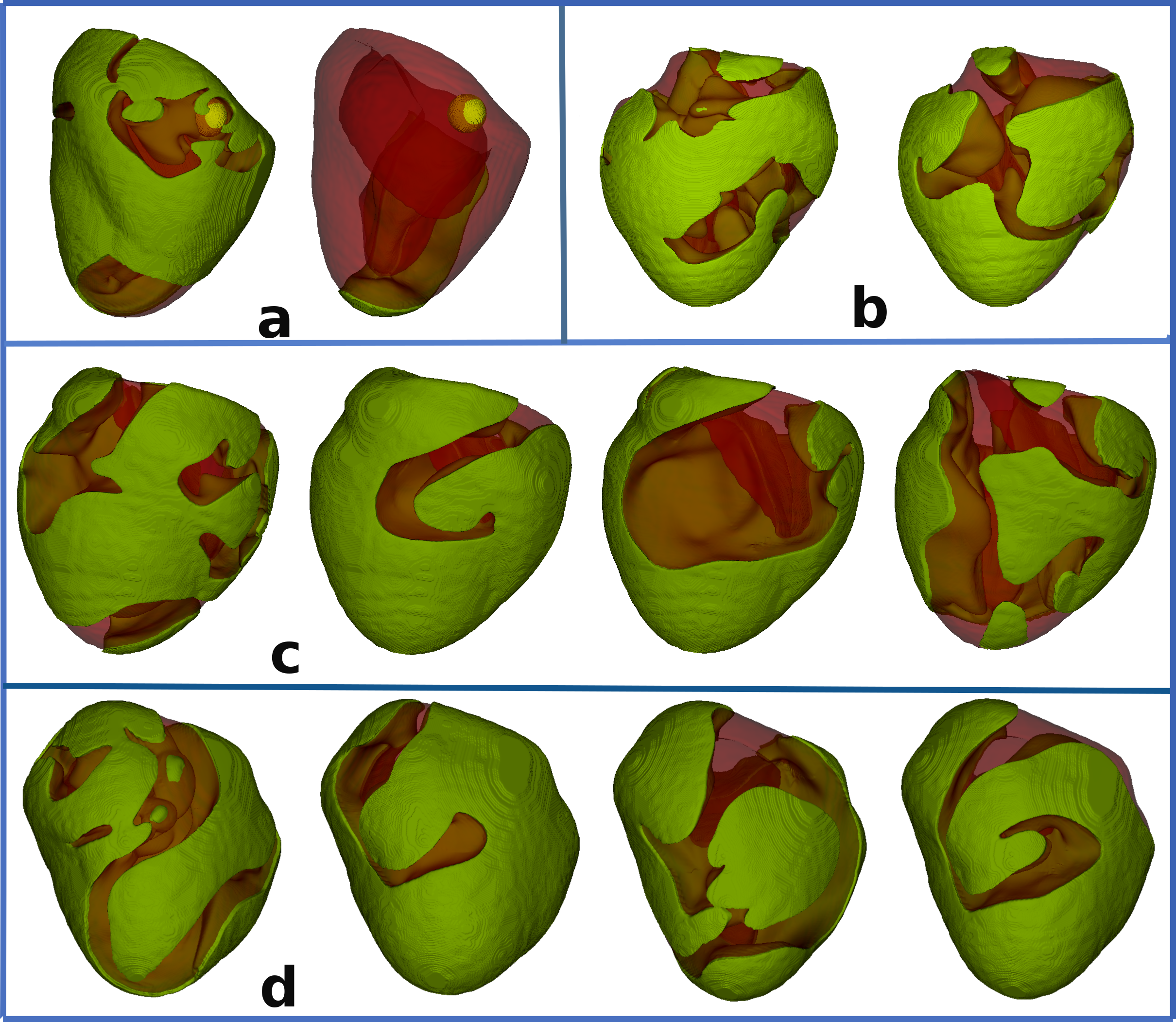}
\end{center}

\caption[Spatiotemporal evolution of scroll waves in the HRD model in the
anatomically realistic canine ventricular geometry, when a medium-sized
spherical inhomogeneity of conduction type is placed at four different
positions and with the initial condition which yields a broken
scroll-wave state.] {Two-level isosurface plots of the transmembrane
potential $V$ illustrating the spatio-temporal evolution of scroll
waves in the HRD model in the anatomically realistic canine ventricular
geometry, when a medium-sized spherical inhomogeneity of conduction
type is placed at the illustrative positions 1 (panel a), 3 (panel b),
2 (panel c), and 4 (panel d) shown in Fig.~\ref{spher_inhomo_pos}. The
initial configuration is obtained from the parameter set that produces
a broken scroll-wave state without the inhomogeneity; we depict the
final configurations here. For the full spatio-temporal evolution of
$V$ see the videos S43 (for panel a), S44 (for panel b), S45 (for panel
c) and S46 (for panel d) in the Supplementary Material~\cite{chapt4movies}.}
\label{cond_spher_1b}
\end{figure}

Figure~\ref{cond_spher_1b} shows the effects of a localized spherical
inhomogeneity (obstacle) of conduction type on a broken-scroll configuration for
four different positions of this obstacle. Panel a shows scroll-wave
dynamics for position 1 of the inhomogeneity; we show isosurfaces of $V$ at an
early time as well as at the final stage: the broken scrolls slowly move away from
the spherical inhomogeneity (depicted in yellow) and finally disappear from the
medium. The panel-c shows the evolution of the scroll wave with the obstacle at
position 2: We see that the broken scrolls have changed to a single rotating
scroll; the scroll breaks up again because of the functional instabilities
introduced by the parameter set we use; the broken scrolls again change to a
rotating scroll; and the last snap-shot depicts another break up. Thus, in this
case, the inhomogeneity prevents scroll break up; it changes broken scrolls
into a single rotating scroll, which again breaks, and this keeps on repeating.
The panel-b shows the spatio-temporal evolution of the scroll wave with the
obstacle at position 3: here, the inhomogeneity cannot suppress the scroll
breakup; in some regions it reduces the extent of break up, but it never
completely changes the broken scrolls into a single scroll; the breakup of
scrolls is considerable in this case. The lowest panel(d) shows the evolution
of a scroll wave for position 4 of the inhomogeneity; in this case the broken
scrolls are slowly changed to a single rotating scroll.

In summary, then, with the localized, medium sized conduction inhomogeneity, for
initial conditions with a rotating scroll, the wave dynamics remains unchanged
for all of the locations of the localized inhomogeneity. If the initial
conditions produce broken scrolls, a localized inhomogeneity drives away the
broken scrolls from the inhomogeneity; and the broken scrolls change into a
rotating scroll, with its tip far away from the inhomogeneity for positions 1,
2, and 4. However, the presence of spherical inhomogeneity at position 3, i.e.,
at the septum, does not affect the broken scroll waves, which continue to
break up. Thus, a medium-sized, spherical conduction inhomogeneity affects the
scroll-wave dynamics slightly differently depending on its location:
specifically, this dynamics depends on the relative distance between the
phase-singularity line of the scroll wave and the position of the inhomogeneity. 
In general, such an inhomogeneity prevents scroll-wave break up, stabilizes broken 
scrolls into a single rotating scroll, and does not affect a single rotating scroll. If,
however, the inhomogeneity is very far from the center of rotation, it cannot
change the broken scrolls into a single rotating scroll. 


We now present the results with small-sized, spherical inhomogeneities
distributed throughout the simulation domain, with three different
concentrations. 

\begin{figure}
\includegraphics[width=0.85\linewidth]{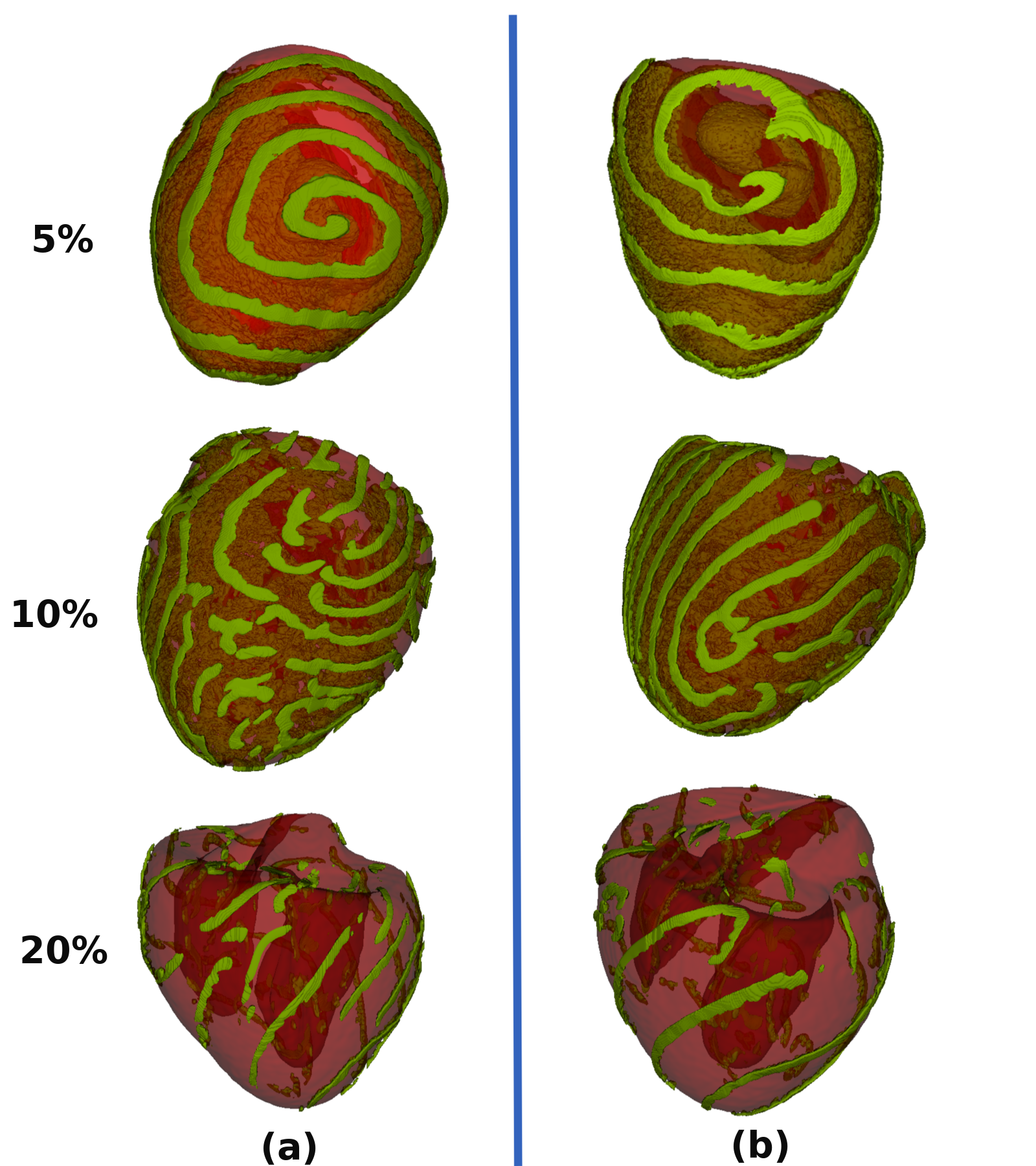}
\caption[Spatiotemporal evolution of scroll waves in the HRD model in the
anatomically realistic geometry for a canine ventricle, with small conduction 
inhomogeneities distributed in the domain, for three different concentrations of the 
inhomogeneities, and with the initial configuration that yields a rotating-scroll 
state on the left and a broken-scroll state on the right.]
{Two-level isosurface plots of the transmembrane potential $V$ illustrating the
scroll-wave dynamics when small conduction inhomogeneities are
distributed in the domain with concentration 5\%, 10\%, and 20\% of the total 
number of sites. The initial configuration is obtained from the parameter set that 
produces a rotating-scroll state (left panel (a)) and a broken-scroll state 
(right panel(b)) a broken scroll-wave state, without the inhomogeneities; we depict the
final configurations here. For the full spatio-temporal evolution of $V$ see the
videos S411, S412, S413 and S414 , S415 and S416 in the Supplementary 
Material~\cite{chapt4movies}.}
\label{cond_distr_5b}
\end{figure}

In Fig.~\ref{cond_distr_5b} we show the spatiotemporal evolution of a scroll
waves when we have distributed inhomogeneities, of the conduction type, for the
following three different concentrations of inhomogeneities: 5\%, 10\%, and
20\%. 

In the left (panel (a)) we depict the spatio-temporal evolution of an initially
rotating scroll wave. The wavelength of the scroll is reduced significantly.
With a 5\% concentration of the inhomogeneities, the wave does not break up; it
continues rotating. For a 10\% concentration of the inhomogeneities, the wave
breaks up and forms a statistically steady broken-scroll state. With a 20\%
concentration of the inhomogeneities, the scroll breaks up to yield a
statistically steady, broken-scroll state; the wavelength is reduced
significantly.

With an initial parameter values which yield broken-scroll configuration,
distributed conduction-type inhomogeneities have a marked effect on scroll-wave
dynamics; in particular, the wavelength of the scroll is significantly reduced.
With a 5\% concentration of these inhomogeneities, the wave does not break up,
but it rotates and curls up more than the initial configuration; in the last
stages of this evolution, we see the two ends of the curled wave interacting
with each other.  With a 10\% concentration of inhomogeneities, the wave breaks
up immediately and we get a statistically steady broken-scroll state.  When we
have a 20\% concentration of the inhomogeneities, the wave breaks up rapidly
and it has a greatly reduced wavelength.  As the percentage of inhomogeneities
increases from 5\% to 20\% the drop in the wavelength and scroll-wave breakup
is substantial.

\begin{table*}[]
\centering
\caption[Summary of our results on the effects of ionic-type inhomogeneities
on scroll-wave dynamics]{This table summarizes our results on the effects of ionic-type inhomogeneities 
on scroll-wave dynamics.}
\begin{tabular}{|c|p{2.5cm}|p{5cm}|p{2cm}|p{5cm}|}
\hline
&\multicolumn{2}{c}{localized spherical inhomogeneity} & \multicolumn{2}{|c|}{distributed small-scale inhomogeneities} \\
\cline{2-5}
\raisebox{-4\normalbaselineskip}[0pt][0pt]{\multirow{4}{*}{\rotatebox[origin=l]{90}{Broken scroll}}}
&position 1 & Broken scroll filaments tend to move away from the inhomogeneity; 
partially; suppresses  scrolls break-up. &{\textbf {5 \%}} & Does not suppress scroll-wave break up.\newline \\
& position 2 & Same dynamics as for position 1. &{\textbf {10 \%} } & Reduces scroll-wave break-up, but does not suppress it completely.\newline \\ 
&position 3 & Does not suppress break up; broken scrolls continues to break up. & {\textbf {20 \% }} & Does not suppress scroll wave break-up.\newline.\\
&position 4 & Same dynamics as for position 3. \newline & & \\ 
\hline
 \raisebox{-1\normalbaselineskip}[0pt][0pt]{\multirow{4}{*}{\rotatebox[origin=b]{90}{Rotating scroll}}} 
&position 1 & Continues rotating. &{\textbf {5 \%}} & Continues rotating; meandering is 
increased.\newline \\
&position 2 & Continues rotating.\newline & {\textbf {10 \%} } & Scroll wave breaks-up occurs. \\
&position 3 & Continues rotating; the scroll tends to move away from the inhomogeneity. 
\newline & {\textbf {20 \% }} & Scroll rotates for a long time without breaking, finally break-up occurs.  \\
&position 4 & Continues rotating. \newline & & \\ 
\hline
\end{tabular}
\label{table2}
\end{table*}
\subsection{Results for Ionic Inhomogeneity}

Before we describe our results in detail, we refer the reader to Table
~\ref{table2}, which summarizes our results for the effects of ionic-type
inhomogeneities on scroll-wave dynamics. We have considered here all of the
cases mentioned in the above section for conduction inhomogeneities.
\begin{figure}
\includegraphics[width=0.95\linewidth]{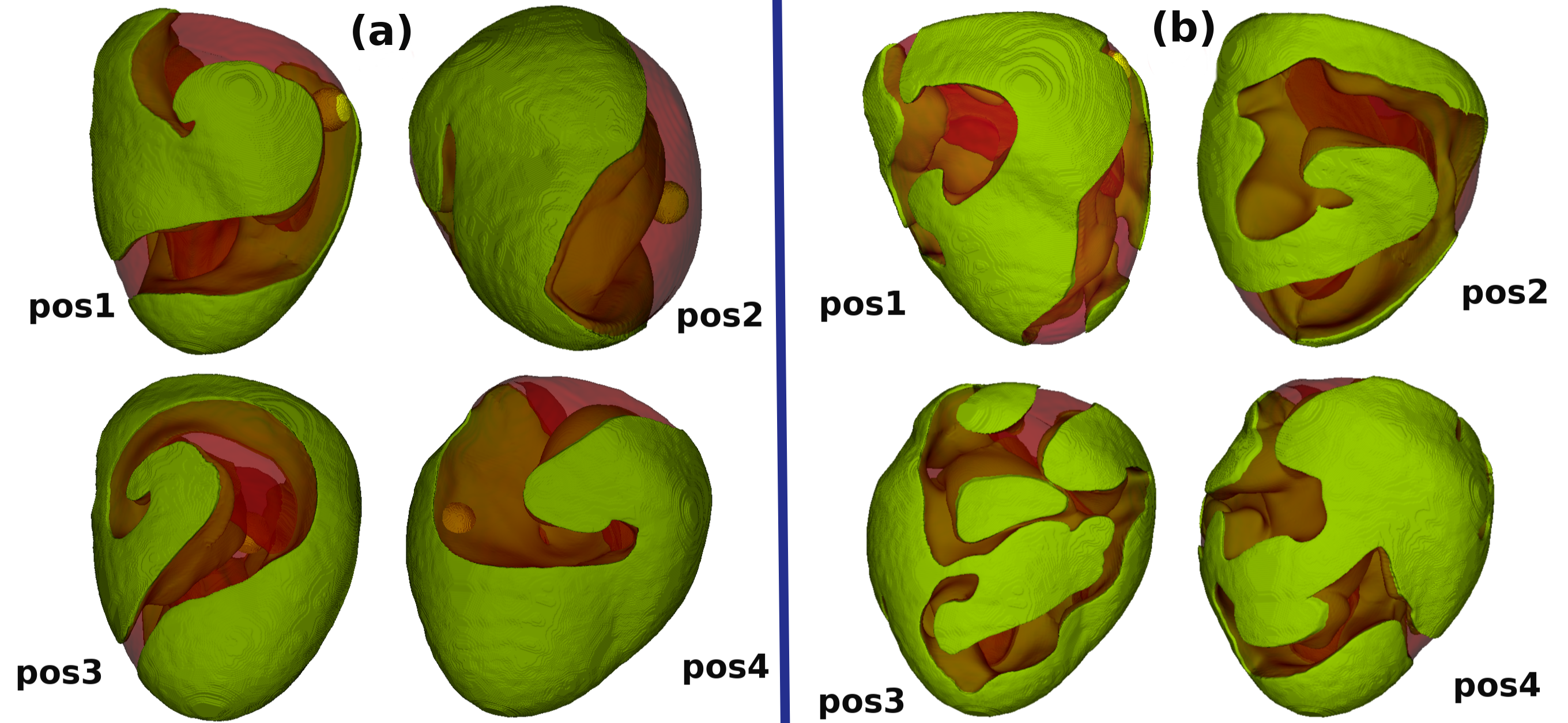}
\caption[Spatiotemporal evolution 
of scroll waves in the HRD model in the anatomically realistic canine
ventricular geometry, with a medium-sized spherical inhomogeneity, of ionic type, 
is placed at four diffrent positions and with the initial configuration that yields a rotating as well as broken scroll-wave state.
]{Two-level isosurface plots of the transmembrane potential $V$ illustrating the
final state of the scroll-wave dynamics when a medium-sized spherical
inhomogeneity, of ionic type, is placed at positions 1 (pos1), 2
(pos2), 3 (pos3), and 4 (pos4), as shown in
Fig.~\ref{spher_inhomo_pos}. The initial configuration is the one that
we obtain from the parameter set that yields  a rotating scroll state
for the left panel(a) and a broken-scroll-wave state for the right
panel(b).  For the full spatio-temporal evolution of $V$ see the videos
S417 , S418 , S419 , and S420 , S421 , S422 , S423 , and S424  in the
Supplementary Material~\cite{chapt4movies}.}

\label{ion_spher_1b}
\end{figure}

In Fig.~\ref{ion_spher_1b}, panel (a) we show the effect of a localized,
spherical inhomogeneity, of ionic type, on scroll-wave dynamics with an initial
rotating-scroll configuration. For this initial condition, we find that the
localized ionic inhomogeneities have no qualitative effect on the rotating
scroll waves. The rotating scroll continues to rotate (Table~\ref{table2}).
For position 3 we see the scroll centre tends to move away from the
inhomogeneity. We conclude, therefore, that scroll-wave dynamics is not
affected substantially by the presence of localized ionic inhomogeneities when
we use rotating-scroll initial conditions. 


Figure~\ref{ion_spher_1b}, panel (b) shows the effect of a localized, spherical
inhomogeneity, of ionic type, on an initial configuration corresponding to
broken-waves.  If the ionic inhomogeneity is at positions 1 or 2, it partially
prevents scroll breakup; and it stabilizes the dynamics to a single rotating
scroll; but, for positions 3 and 4, the ionic inhomogeneity does not suppress
the break up. Also, an ionic inhomogeneity does not completely remove the
broken scrolls from the domain, as we have observed for position 1 for a
conduction-type inhomogeneity.

\begin{figure}
\includegraphics[width=0.85\linewidth]{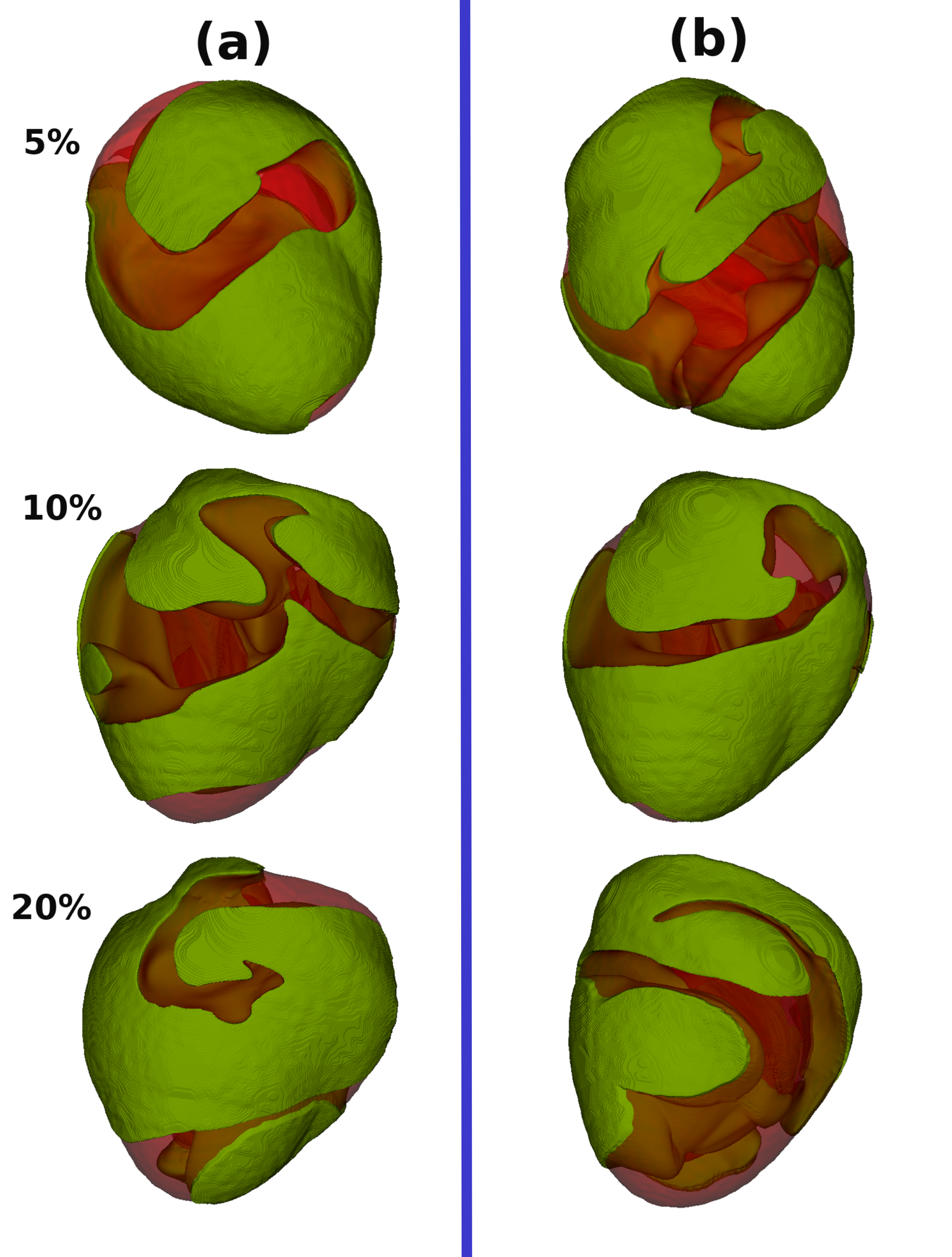}

\caption[Spatiotemporal evolution of scroll waves in the HRD model in the
anatomically realistic canine ventricular geometry, with small ionic
inhomogeneities distributed in the domain, with three different
concentrations, and with the initial configurations that yield a
rotating scroll wave as well as wave breaking.] {Two-level isosurface
plots of the transmembrane potential $V$ illustrating scroll-wave
dynamics, when small ionic inhomogeneities are distributed in the
domain with a concentration 5\%, 10\% and 20\% of the total number of
sites. The initial configuration corresponds to the parameter set that
yields, in the absence of inhomogeneities, a rotating  scroll wave
[panel (a)], and scroll wave break-up  [panel(b)].  For the full
spatio-temporal evolution of $V$ see the videos S425 , S426, S427,
S428, S429, and S430 in the Supplementary Material~\cite{chapt4movies}.}

\label{ion_distr_5r}
\end{figure}

We now examine the effects of distributed inhomogeneities, of ionic type, on
two initial configurations in Fig.~\ref{ion_distr_5r}; for each case, we depict
isosurfaces of $V$ from the last stage of our simulation.  For small-sized,
distributed inhomogeneities, with a rotating-scroll initial configuration, we
see that (as in the case of conduction inhomogeneities), for small
concentrations of inhomogeneities (5\%), the rotating scroll wave is unaffected
by them; but, as this concentration increases, the inhomogeneities cause the
rotating scroll to break up. The break up is not as fast, intense, or
statistically steady, as in the case of conduction inhomogeneities.
Furthermore, here the wavelength of the scrolls does not change (as it does for
conduction inhomogeneities).  We observe that the extent of such break up is more
for 10\% inhomogeneities than for 20\%. 


Figure.~\ref{ion_distr_5r}, panel (b), shows the effects on scroll-wave
dynamics, of distributed inhomogeneities of ionic type, for an initial
broken-scroll configuration. We observe that, for 10\% distributed
inhomogeneities, the scroll break up is partially suppressed. In other cases
the such break up is not suppressed. 

We conclude that the effect of ionic-type inhomogeneities on scroll-wave
dynamics of broken scrolls are different, in detail, from those of
conduction-type of inhomogeneities. For distributed, small-sized
inhomogeneities, conduction inhomogeneities break up the scrolls significantly;
ionic inhomogeneities prevent scroll-wave break up for moderate(10\%)
concentration of inhomogeneities.  Localized inhomogeneities have almost the
same effect on scroll-wave dynamics as conduction-type obstacles. At two
positions (1 and 2), they prevent the breaking up of scrolls by partially
changing broken waves to rotating scrolls. For ionic inhomogeneities at
positions 3 and 4, we do not see this behavior. Thus, scroll-wave dynamics
also depends on the position of the inhomogeneity. Finally, we note that a
localized ionic inhomogeneity does not have a significant effect on an
initially rotating scroll wave as in the case of localized conduction
inhomogeneity. 

\section{Conclusions}
\label{sec:Conclusions}

We have carried out a detailed \textit{in silico} study to examine the effects
of different types of inhomogeneities on scroll-wave dynamics in dog
ventricles. Our work goes well beyond earlier studies because we have used an
anatomically realistic model for canine ventricles, namely, the HRD model; and we
have also included muscle-fiber orientation.  Our study has elucidated the
effects of \textit{conduction} and \textit{ionic} inhomogeneities on scroll
waves in this model. 

In particular, we have found that small-scale, distributed, conduction
inhomogeneities have the most marked effect on scroll-wave dynamics: these
inhomogeneities reduce the wavelength significantly at all concentrations.  Low
concentration of the inhomogeneities ($\simeq$ 5\%) do not break up the scroll
wave; but higher concentrations(10\% and above) do break up the rotating scroll
and, indeed, break up the broken scrolls too.  A localized, large-scale
conduction inhomogeneity, partially suppresses wave break ups and stabilizes
broken waves to form a rotating wave (for most positions of the inhomogeneity);
and it does not affect the break up of scrolls in one position (in the septum).
However, if the medium already has a stable rotating wave, the presence of such
inhomogeneities does not affect the wave dynamics. 

We find, for the ionic inhomogeneities that we have studied, at medium(10\%)
concentrations of distributed, small-scale inhomogeneities partially suppress
scroll-wave break up; at other concentrations, the such inhomogeneities do not
affect the broken waves.  Higher (10\% and above) concentrations of distributed
inhomogeneities, can cause  the partial break up of initially rotating scroll
waves; and low concentrations do not affect the scroll-wave dynamics. A
localized, large-scale, ionic-type inhomogeneity stabilizes the broken-wave
dynamics at two positions; and it does not affect the dynamics when it is at
the other positions (see above).  The dynamics of a stable rotating wave is not
affected significantly by localized ionic inhomogeneities.  The scrolls tend to
move away from local inhomogeneities of both conduction and ionic types.  When
the position of the inhomogeneity is away from the region of the phase
singularity of the scrolls, it does not suppress the break up significantly. 

In summary, our study shows that localized conduction inhomogeneities can
suppress scroll-wave break up and chaos, to some extent; and distributed,
conduction inhomogeneities can lead to a statistically steady broken-wave state
with spatiotemporal chaos.  Stable rotating scrolls in the medium are not
affected by localized inhomogeneities, but are broken by distributed
inhomogeneities at high concentrations, for both conduction and ionic types
of inhomogeneities. Ionic type inhomogeneities can suppress scroll-wave
break up.

Our results can be compared fruitfully with a similar \textit{in silico}
study~\cite{rpm-inhomo} of scroll-wave dynamics in a mathematical model for a
pig heart. This study also finds that small-scale, distributed, conduction
inhomogeneities have the most marked effect on scroll-wave dynamics, as we have
found for the HRD canine-ventricular model.  The main findings of the above
study are conduction-type inhomogeneities become increasingly important, in the
case of multiple, randomly distributed, obstacles at the cellular scale ($0.2$ mm
$- 0.4$ mm). Such configurations can lead to scroll-wave break up; and  ionic
inhomogeneities affect scroll-wave dynamics significantly at large length
scales, when these inhomogeneities are localized in space at the tissue level
($5$ mm $- 10$ mm). In such configurations, these inhomogeneities can (a) attract
scroll waves, by pinning them to the heterogeneity, or (b) lead to scroll-wave
breakup.  The authors of Ref.~\cite{rpm-inhomo} used an initial configuration with a 
stable rotating spiral. They found no qualitative change in the dynamics of such a 
wave when small-scale ionic distributed inhomogeneities were introduced and
also when large, localized conduction inhomogeneities were included.

We also find that large, localized, conduction inhomogeneities do not affect a
stable rotating wave. Thus, for such inhomogeneities, our results match with
those for this pig-heart study~\cite{rpm-inhomo}, with the same initial
conditions. We study broken-wave initial conditions too.  For ionic
inhomogeneities, our HRD-model studies show that the stable rotating wave
finally breaks up for large concentration of distributed inhomogeneities; this
is in contrast to pig-heart study of Ref.~\cite{rpm-inhomo}, in which such a
wave is not significantly affected by these inhomogeneities.  Conduction
inhomogeneities affect the diffusion part of the governing equations; this is
the same for all models, so the results also turn out to be the same for
canine- and porcine-ventricular models.  The reaction part of the equations is
model specific; and the ionic inhomogeneities that are used in these two
studies also have different conductances; thus, the results, with such
inhomogeneities, are model specific.

We mention some limitations of our study. We have used a monodomain model for
the cardiac-tissue equations in our study; bidomain models are more realistic
than monodomain ones; however, a recent study~\cite{Potse} has shown that the
latter are adequate when currents are low, as in our study. To impose boundary
conditions we have used a phase-field approach; this can also be done with a
finite-element model, which we leave for future work.  The inhomogeneities
could be handled in a more realistic way than in our study in terms of shape,
composition, etc. 

We hope our work will lead to other detailed studies of scroll-wave dynamics in
realistic models for mammalian hearts. Only by contrasting the results of such
studies can we develop a detailed understanding of VT and VF, which is a
prerequisite for the efficient elimination (i.e., defibrillation) of such
life-threatening cardiac arrhythmias.
\section{Acknowledgments}
We thank  Council of Scientific and Industrial Research, University Grants
Commission and  Department of Science and Technology for support, and
Supercomputing Education and Research Centre (IISc) for computational
resources.

\end{document}